\documentclass[twocolumn,preprintnumbers,amssymb,amsmath,9pt,superscriptaddress]{revtex4}[10pt]
\usepackage{graphicx}
\usepackage{dcolumn}
\usepackage{bm}
\begin{document} 
\input epsf.tex
\newcommand{\beq}{\begin{eqnarray}}
\newcommand{\eeq}{\end{eqnarray}}
\newcommand{\nn}{\nonumber}
\def\ltap{\ \raise.3ex\hbox{$<$\kern-.75em\lower1ex\hbox{$\sim$}}\ }
\def\gtap{\ \raise.3ex\hbox{$>$\kern-.75em\lower1ex\hbox{$\sim$}}\ }
\def\CO{{\cal O}}
\def\CL{{\cal L}}
\def\CM{{\cal M}}
\def\tr{{\rm\ Tr}}
\def\CO{{\cal O}}
\def\CL{{\cal L}}
\def\CM{{\cal M}}
\def\mpl{M_{\rm Pl}}
\newcommand{\bel}[1]{\be\label{#1}}
\def\al{\alpha}
\def\bt{\beta}
\def\eps{\epsilon}
\def\eg{{\it e.g.}}
\def\ie{{\it i.e.}}
\def\mn{{\mu\nu}}
\newcommand{\rep}[1]{{\bf #1}}
\def\be{\begin{equation}}
\def\ee{\end{equation}}
\def\bea{\begin{eqnarray}}
\def\eea{\end{eqnarray}}
\newcommand{\eref}[1]{(\ref{#1})}
\newcommand{\Eref}[1]{Eq.~(\ref{#1})}
\newcommand{\gsim}{ \mathop{}_{\textstyle \sim}^{\textstyle >} }
\newcommand{\lsim}{ \mathop{}_{\textstyle \sim}^{\textstyle <} }
\newcommand{\vev}[1]{ \left\langle {#1} \right\rangle }
\newcommand{\bra}[1]{ \langle {#1} | }
\newcommand{\ket}[1]{ | {#1} \rangle }
\newcommand{\ev}{{\rm eV}}
\newcommand{\kev}{{\rm keV}}
\newcommand{\Mev}{{\rm MeV}}
\newcommand{\gev}{{\rm GeV}}
\newcommand{\tev}{{\rm TeV}}
\newcommand{\mev}{{\rm MeV}}
\newcommand{\mnu}{\ensuremath{m_\nu}}
\newcommand{\mlr}{\ensuremath{m_{lr}}}
\newcommand{\acc}{\ensuremath{{\cal A}}}
\newcommand{\mav}{MaVaNs}
\newcommand{\disc}[1]{{\bf #1}}

\title{Supersplit Supersymmetry}
\author{Patrick J. Fox}\affiliation{Santa Cruz Institute for Particle Physics, Santa Cruz, CA,  95064}
\author{David E. Kaplan}\affiliation{Dept. of Physics and Astronomy,
Johns Hopkins University, Baltimore, MD  21218}
\author{Emanuel Katz}\affiliation{Stanford Linear Accelerator Center, 2575 Sand Hill Rd.
Menlo Park, CA 94309}\affiliation{Dept. of Physics, Boston University, Boston, MA  02215}
\author{Erich Poppitz}\affiliation{Department of Physics, University of Toronto, 60 St George St, 
Toronto, ON M5S 1A7, Canada}
\author{Veronica Sanz}\affiliation{Universitat de Granada, Campus de
  Fuentenueva, Granada, Spain}
\author{Martin Schmaltz}\affiliation{Dept. of Physics, Boston University, Boston, MA  02215}
\author{Matthew D. Schwartz}\affiliation{University of California, Dept. of Physics,
Berkeley, CA 94720-7300}
\author{Neal Weiner}\affiliation{Center for Cosmology and Particle Physics,
  Dept. of Physics, New York University,
New York , NY 10003}

\date{April 1, 2005}
\begin{abstract}
The possible existence of an exponentially large number of vacua in string
theory behooves one to consider possibilities beyond our traditional notions
of naturalness.  Such an approach to electroweak physics was recently used in
``Split Supersymmetry", a model which shares some successes and cures some
ills of traditional weak-scale supersymmetry by raising the masses of scalar
superpartners significantly above a TeV.  Here we suggest an extension - we
raise, in addition to the scalars, the gaugino and higgsino masses to much
higher scales.  In addition to maintaining many of the successes of Split
Supersymmetry - electroweak precision, flavor-changing neutral currents and
CP violation, dimension-4 and 5 proton decay - the model also allows for
natural Planck-scale supersymmetry breaking, solves the gluino-decay problem,
and resolves the coincidence problem with respect to gaugino and Higgs
masses.  The lack of unification of couplings suggests a natural solution to
possible problems from dimension-6 proton decay.  While this model has no
weak-scale dark matter candidate, a Peccei-Quinn axion or small black holes
can be consistently incorporated in this framework.
\end{abstract}
\maketitle
\section{Introduction}
For a many years, our motivations for considering new physics at the weak scale have been strongly influenced by ideas of naturalness. The difficulty of maintaining light scalars in a theory with a high cutoff has led us to consider compositeness, supersymmetry, or pseudo-goldstone boson theories at the weak scale.

Recently, it has been realized that the broad string landscape may have an exponentially large number of metastable vacua \cite{Kachru:2003aw,Susskind:2003kw, Douglas:2003um}.  With so many vacua, it is possible to appeal to Weinberg's argument for a solution to the cosmological constant problem based upon a scan over many possible universes \cite{Weinberg:1987dv}. 

Of course, the presence of such a severe fine-tuning may involve scannings and associated tunings of other parameters. The case that this may impact our expectations of the weak scale, and in particular supersymmetric theories was made by Arkani-Hamed and Dimopoulos \cite{Arkani-Hamed:2004fb}.  In ``Split Supersymmetry'', all but one scalar of the many of the minimal supersymmetric standard model (MSSM) are given very large masses. Of the two scalar superpartners of the Higgsinos, one linear combination remains light and then acquires a vacuum expectation value (vev) which breaks electroweak symmetry and gives masses to the weak gauge bosons. The fermion masses, which can be protected by symmetries, remain small.

While flying in the face of naturalness, unification of gauge couplings
\cite{Dimopoulos:1981zb, Dimopoulos:1981yj} and weak scale dark matter - both often used to compel
supersymmetry - are maintained. This makes it a phenomenologically appealing model, even if the notions of what constitutes a natural point in the string landscape are still being worked out \cite{Douglas:2003um, Dine:2005yq}.


It is exciting to consider modifications to this model. However, unlike
traditional unwieldy model-building, in which additional fields are added and
their phenomenological consequences studied, here we remove fields and their
associated phenomenological problems. This has already been proposed in a
limited form in \cite{Arkani-Hamed:2005yv} where the gauginos were decoupled
from the weak scale in addition to the scalars (alternatively,
the higgsinos could be decoupled \cite{Cheung:2005ba}).
Here a dark matter candidate remains in the Higgsinos, but gauge coupling unification occurs at a low scale ($10^{14} \gev$), which would typically induce unacceptable rates of proton decay.

\section{The Model}
The next logical extension would be to decouple one Higgsino, in addition to the scalars and gauginos. Unfortunately, due to anomalies, this is not possible, so we take the next simplest possibility, which is to decouple both Higgsinos. The low energy effective theory consists of $SU(3)\times SU(2) \times U(1)$ gauge fields and three generations of quarks and leptons, as well as one scalar (whose mass is tuned to be light) which is responsible for electroweak symmetry breaking. 
The schematic of this model in comparison to traditional and Split SUSY is presented in figure 1.

 The Lagrangian for this model is simply
 \bea
 {\cal L} &=& -\frac{1}{4 g^2_1}W_{\mu\nu} W^{\mu\nu} - \frac{1}{4
 g^2_2}B_{\mu\nu} B^{\mu\nu} -\frac{1}{4 g^2_s}G_{\mu\nu} G^{\mu\nu}
\nonumber \\ &+&\bar
 \psi_f (i\gamma^\mu D_\mu - m_f) \psi_f + D_\mu h D^\mu h^* - V(h)
 \eea
where $f$ indexes the various fermions, and $D$ is the appropriate covariant derivative.

\begin{figure}
\epsfxsize=3.0 in \epsfbox{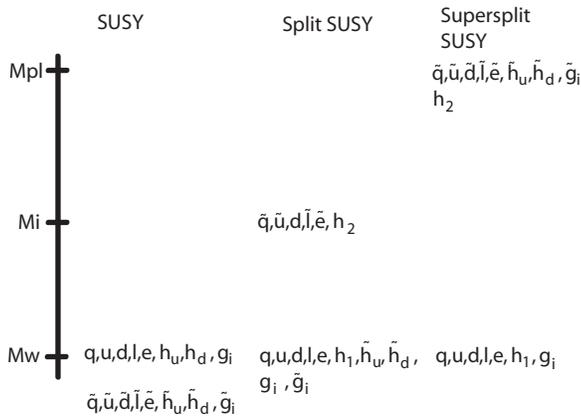} 
\caption{\label{fig:epsart} Mass scales in the MSSM, Split SUSY and Supersplit SUSY.}
\end{figure}

\section{Phenomenology}
A few comments are immediately in order: first, because the scalars, gauginos, and Higgsinos are decoupled we needn't worry about flavor changing neutral currrents or EDMs. Because most states are decoupled to the Planck scale \cite{cites}, baryon number violating interactions are suppressed at a safe level. 

Unification is predicted {\it not} to happen at $10^{15} - 10^{16}$ GeV. Indeed, the absence of traditional gauge coupling unification in this model is a significant strength of it over most SUSY models. Such unification compels one to consider unified models which are often disastrous both from a phenomenological perspective, due to proton decay, and aesthetically, due to the numerous theoretical problems, such as the doublet-triplet splitting problem, light fermion masses, etc. Here, we are freed from these issues and the model is conceptually much simpler.

Remarkably, however, our model is consistent with unification {\it at the
  Planck scale}! 
Assuming a ratio $M_{GUT}/M_{Planck} \sim 1/2$, the three
gauge couplings unify at the new GUT scale with up to 50\% theoretical uncertainty coming
from higher dimensional operators such as
\begin{equation}
\frac{1}{g^2} \frac{\tr{F_{\mu\nu}F^{\mu\nu}\Sigma}}{M_{Planck}} \ .
\end{equation}
Thus the dream of gauge-gravity unification can be realized without resorting to extra dimensions (or other new physics) below the Planck scale.

Corrections to electroweak observables are small, and are compatible with a Higgs boson lighter than $280\ \gev$ \cite{Eidelman:2004wy}, which is still above the limits from collider searches.

The absence of new colored states should make this model simple to
distinguish from both SUSY and Split SUSY, although a challenge to
distinguish from the model with Higgsinos, making the ILC essential. 

While the model predicts no weak scale dark matter (see figure 1),
a Peccei-Quinn axion or dark matter in the form of small-mass black
holes with their associated rich phenomenology \cite{rich1, rich2}
could be accomodated
naturally.

\section{Conclusions}
We have presented what we believe to be the simplest known supersymmetric
theory, consistent with available data. Indeed, it may be the simplest theory
which agrees with the data, whether supersymmetric or not, at least in terms
of the light field content. In future work, we plan to compare the phenomenological success of this model to all other models.

Remarkably, it is only in the modern context of the landscape that we can
appreciate such a finely tuned theory. It would have been rejected out of hand by traditional effective field theorists only a decade ago. In the modern context, it is a strong competitor to other theories of physics at the weak scale. 

An open question that remains is the origin of vacuum selection on the
landscape, an issue too great to discuss here. While the so-called ``atomic
principle" \cite{Agrawal:1997gf} has been used as part of an anthropic
argument for fine tuning of the weak scale, a completely independent (and
predictive) argument can be made without resorting to changes in weak-scale
physics.  Here however we make only a qualitative statement \cite{quant}.
One could argue that the existence of fine tuning in nature dramatically
increases the time scale over which fundamental physical laws are discovered.
Rapid discovery of fundamental laws necessarily advance the discovery of
weapon-systems with a global impact.  In such a universe without fine-tuning,
theoretical physics may not be possible.

Regardless, the future of physics will no doubt shed much light on these and other exciting questions.

\vskip 0.2in
\noindent {\bf Note added:} While this work was being completed, we became aware of \cite{glashow,salam,weinberg}, a series of conference talks where a similar model was considered. While there are some similarities (specifically, field content and interactions), the philosophy is completely unrelated.

\vskip 0.2in
\noindent{\bf Acknowledgements:}  We thank S. Alon and Y. Banaji for their hospitality when this work was initiated. 


 %


\begin{thebibliography}{99}
\bibitem{cites}
We should point out that while we are focusing on the case where these fields are at the Planck scale, in some string theories \cite{Horava:1995qa} it may be natural to have them at a lower scale, such as the GUT scale.
\end{thebibliography}
\end{document}